\documentclass[%
reprint,
amsmath,amssymb,
aps,
]{revtex4-2}

\usepackage{graphicx}
\usepackage{dcolumn}
\usepackage{bm}
\usepackage{xcolor}
\usepackage{amsmath}

\newcommand{\cU}{{\mathcal U}}

\newcommand{\CO}{{\mathcal O}}
\newcommand{\CR}{{\mathcal R}}
\newcommand{\CC}{{\mathcal C}}

\newcommand{\CV}{{\mathcal V}}
\newcommand{\CM}{{\mathcal M}}
\newcommand{\CB}{{\mathcal B}}
\newcommand{\CZ}{{\mathcal Z}}

\newcommand{\CS}{{\mathcal S}}
\newcommand{\CL}{{\mathcal L}}
\newcommand{\CW}{{\mathcal W}}

\newcommand{\cH}{{\mathcal H}}
\newcommand{\CH}{{\mathcal H}}
\newcommand{\tq}{{\mathtt q}}
\newcommand{\ta}{{\mathtt a}}
\newcommand{\tb}{{\mathtt b}}

\newcommand{\1}{{1}}
\newcommand{\2}{{2}}
\newcommand{\3}{{3}}

\begin{document}

\preprint{TIFR/TH/22-34}

\title{A new multi-partite entanglement measure and its holographic dual}

\author{Abhijit Gadde}
\author{Vineeth Krishna}
\author{Trakshu Sharma}
\affiliation{$\quad$ \\Department of Theoretical Physics,\\ Tata Institute of Fundamental Research}

\begin{abstract}
In this letter we define a natural generalization of the von Neumann entropy to multiple parties that is symmetric with respect to all the parties. We call this measure multi-entropy. We show that for conformal field theories with holographic duals, the multi-entropy is computed by the area of an appropriate ``soap-film'' anchored on the boundary. We conjecture the quantum version of this prescription that takes into account the sub-leading corrections in $G_N$. 
\end{abstract}


\maketitle

\section{\label{sec:level1}Introduction}
In recent years, quantum information theoretic notions such as entanglement entropy \cite{Calabrese:2004eu, Ryu:2006bv, Hubeny:2007xt, Casini:2011kv, Dong:2013qoa, Engelhardt:2014gca} have been immensely useful in shedding light on some of the important questions pertaining to quantum gravity . 
In particular, it has helped understand  holographic encoding of the gravitational Hilbert space and in turn, the Black-hole entropy \cite{Bekenstein:1973ur, Hawking:1975vcx} and the  information paradox. However, with the exception of reflected entropy \cite{Dutta:2019gen} and negativity \cite{Calabrese:2012ew, Kusuki:2019zsp}, almost all the discussion has been centered on the entanglement between two parties i.e. on bi-partite entanglement. It is expected that knowing multi-partite entanglement structure in quantum gravity would refine our understanding of the  holographic encoding of gravitational Hilbert space in conformal field theory Hilbert space. To this end, we introduce a multi-partite entanglement measure and its holographic dual.

\section{The Measure}

Let $|\Psi\rangle\in \otimes_{\ta=\1}^\tq {\cal H}_\ta$ be a quantum state of a $\tq$-party system. Let ${\rm dim}\,\CH_\ta=d_\ta$ and $|\alpha_\ta\rangle, \, \alpha_\ta\in \{1,\ldots, d_\ta\}$ be its  orthonormal basis.  In $|\alpha_\ta\rangle$ basis, the state $|\Psi\rangle$ is given as 
\begin{align}
    |\Psi\rangle=\sum_{\alpha_\1=1}^{d_1}\ldots\sum_{\alpha_\tq=1}^{d_\tq} \,\psi_{\alpha_\1\ldots \alpha_\tq} |\alpha_\1\rangle\otimes\ldots\otimes|\alpha_\tq\rangle.
\end{align}
A $\tq$-party entanglement measure is the information contained in the wavefunction $\psi$ (and its complex conjugate) that is invariant under ``local unitary transformations''. They are products of unitary operations $\cU_\ta\in U(d_\ta)$ performed in individual system $\ta$. Naturally, such a measure is obtained by contracting $\alpha_\ta$ indices for all $\ta$. Now $\alpha_\ta$ is a fundamental index on $\psi$ and an anti-fundamental index of $\bar \psi$. Hence, in order to construct invariants we need an equal number of $\psi$'s and $\bar \psi$'s. We will call this number the replica number $n$. We index the replicas by the superscript $(i)$. In particular, the Hilbert space $\cH_a$ of the $i$-th replica is denoted as $\cH_\ta^{(i)}$ and the associated basis as $|\alpha_\ta^{(i)}\rangle$. The wavefunction of the $i$-th replica is then $\psi_{\alpha_\1^{(i)}\ldots\alpha_\tq^{(i)}}$ and similarly its conjugate is $\bar \psi^{\alpha_\1^{(i)}\ldots \alpha_\tq^{(i)}}$.

\subsection{Bi-partite}

Let us consider the following bi-partite entanglement measure  with a fixed replica number $n$
\begin{align}\label{renyi}
    \CZ_n\equiv {\rm Tr}(\rho^n), \quad {\rm where}\quad \rho_{\alpha^{(1)}_\1}^{\alpha^{(2)}_\1}\equiv \psi_{\alpha^{(1)}_\1\alpha^{(1)}_\2}\bar \psi^{\alpha^{(2)}_\1 \alpha^{(1)}_\2}.
\end{align}
Here $\rho$ is the density matrix for party $\1$. With this notation, the squared norm of $|\Psi\rangle$ is $\CZ_1$. This measure is related to the  familiar $n$-th Renyi entropy $S_n$ as 
\begin{align}
    S_n=\frac{1}{1-n}{\rm Log} (\CZ_n/\CZ_1^n).
\end{align} 
Let us reformulate this measure using permutations acting on replicas. This will be useful while dealing with entanglement in quantum field theories using twist operators.
\begin{align}
    \CZ_n=(\psi_{\alpha^{(1)}_\1\alpha^{(1)}_2}\psi_{\alpha^{(2)}_1\alpha^{(2)}_\2}\ldots)(\bar \psi^{\alpha^{(1)}_\1\alpha^{(\sigma\cdot 1)}_\2}\bar \psi^{\alpha^{(2)}_\1\alpha^{(\sigma\cdot 2)}_\2}\ldots )
\end{align}
where $\sigma$ is a cyclic permutation element acting in the space of replicas $i=1,\ldots,n$ of the second party i.e. $\sigma(\alpha_\2^{(i)})=\alpha_\2^{(i+1)}$ with $\alpha_\2^{(n+1)}\equiv \alpha_\2^{(1)}$. Because this permutation acts only on the replica copies of party-$\2$, we denote it as $\sigma_\2$.
A priori, it would seem that we should specify two  permutation elements $(\sigma_\1,\sigma_\2)$, the first one acting on replicas of party-$\1$ and the other acting on replicas of party-$\2$ to get the most general index contraction, i.e.
\begin{align}\label{2-measure}
    (\psi_{\alpha^{(1)}_\1\alpha^{(1)}_2}\psi_{\alpha^{(2)}_1\alpha^{(2)}_\2}\ldots)(\bar \psi^{\alpha^{(\sigma_\1\cdot 1)}_\1\alpha^{(\sigma_\2\cdot 1)}_\2} \bar \psi^{\alpha^{(\sigma_\1\cdot 2)}_\1\alpha^{(\sigma_\2\cdot 2)}_\2}\ldots )
\end{align} 
However, we can always relabel replicas of $\bar \psi$ so that one of the $\sigma$'s, say $\sigma_\1$, is brought to ${\rm id}$ form. In other words, $(\sigma_\1,\sigma_\2)$ and $(\sigma_\1 g, \sigma_\2 g), g\in {\mathbb S}_n$ yield the same measure. We express this as an equivalence relation
\begin{align}\label{2-right}
    (\sigma_\1,\sigma_\2)\sim (\sigma_\1,\sigma_\2)\, g
\end{align}
There is yet another equivalence, namely under simultaneous relabeling of replica copies of \emph{both} $\psi$'s and $\bar \psi$'s. Such a relabeling leads to conjugation of both $\sigma_\ta$ by some $h\in {\mathbb S}_n$. We express this as the equivalence relation
\begin{align}\label{2-conjugate}
    (\sigma_\1,\sigma_\2)\sim h^{-1}\,(\sigma_\1,\sigma_\2)\, h.
\end{align} 
If we use equation \eqref{2-right} to set $\sigma_\1={\rm id}$, then thanks to equation \eqref{2-conjugate}, the bi-partite entanglement measure depends only on the conjugacy class of $\sigma_\2$. Conjugacy class in ${\mathbb S}_n$ is specified by cycle decomposition $\{p_k\}$ where $p_k$ is the number of $k$-cycles such that $\sum_k k\,p_k=n$. For this class, it is easy to see that the resulting measure \eqref{2-measure} is 
\begin{align}
    \prod_k\,({\rm Tr}(\rho^k))^{p_k}=\prod_k\,(\CZ_k)^{p_k}.
\end{align}
This shows that the ring of bi-partite measures is generated by $\CZ_k$'s.

\subsection{Multi-partite}

Following this discussion, we can write a general multi-partite measure by giving permutation elements $(\sigma_\1,\ldots,\sigma_\tq)$ with equivalence relations
\begin{align}
    (\sigma_\1,\ldots,\sigma_\tq)&\sim (\sigma_\1,\ldots,\sigma_\tq)\cdot g \label{right}\\
    (\sigma_\1,\ldots,\sigma_\tq)&\sim h^{-1}\,(\sigma_\1,\ldots,\sigma_\tq)\,h.\label{conjugate}
\end{align}
Unlike the bi-partite case, where Renyi entropies are the only independent measures of entanglement, for the case of three or higher number of parties, the number of measures increase exponentially or faster with the replica number $n$. In this letter, we will not concern ourselves with general $\tq$-party measures but rather focus on a particular family of measures labeled by an integer that is symmetric under the exchange of all the parties.

As in the bi-partite case, equation \eqref{right} can be used to set $\sigma_\1={\rm id}$. But this ``gauge fixing'' obscures the symmetry between all the parties as it treats party-$\1$ differently from others. So we will not use the gauge freedom \eqref{right} and \eqref{conjugate} just yet and specify all $\sigma_\ta$'s. We will do so with $n^{\tq}$ replicas. We index the replicas with a $\tq$-dimensional index vector $(i_\1,\ldots,i_\tq)$ where each $i_\ta=\{1,\ldots,n\}$. 
The permutation element $\sigma_\ta$ is defined as the cyclic element acting only on $i_\ta$: 
\begin{align}\label{sigma-def}
    \sigma_\ta\cdot (\ldots,i_\ta,\ldots)=(\ldots,i_\ta+1,\ldots),\quad \ta=1,\ldots,\tq.
\end{align}
As an element of the permutation group ${\mathbb S}_{n^{\tq}}$, its conjugacy class is  $p_{n}=n^{\tq-1}$ with all other $p_k=0$.  It is also clear that this measure is symmetric in all the $\tq$ parties. 

Now we use the gauge freedom \eqref{right} with $g=\sigma_\1^{-1}$. This gives us an equivalent  set of permutation elements 
\begin{align}
    ({\rm id},\sigma_\2\sigma_\1^{-1},\ldots,\sigma_\tq \sigma_\1^{-1}).
\end{align}
Note that the action of any of these permutations on $(i_\1,\ldots,i_\tq)$ keeps the sum $\sum_\ta\,i_\ta$ invariant. As a result, their action on  $n^{\tq}$ replicas splits into $n$ orbits of $n^{\tq-1}$ elements with $\sum_\ta i_\ta={\rm constant} ({\rm mod}\, n)$. Each orbit gives rise to the same invariant of $\psi$'s and $\bar\psi$'s. 
It is convenient to work with a single orbit, say with $\sum_\ta i_\ta=0({\rm mod}\, n)$. For this orbit, we can use the relabeling freedom \eqref{conjugate} to set the replica index $i_\1$ to $1$. With this gauge fixing,  we get a convenient presentation of the rest of the permutation elements $(\sigma_\2\sigma_\1^{-1},\ldots,\sigma_\tq\sigma_\1^{-1})\equiv (\hat \sigma_\2,\ldots,\hat \sigma_\tq)$.
\begin{align}\label{sigma-hat}
    \hat \sigma_\ta\cdot (\ldots,i_\ta,\ldots)=(\ldots,i_\ta+1,\ldots),\quad \ta=\2,\ldots,\tq.
\end{align}  
We have denoted the gauge fixed permutation elements as $\hat \sigma_\ta$ to avoid confusion with un-gauge fixed permutation elements $\sigma_\ta$.
The index set in equation \eqref{sigma-hat} starts from $i_\2$ because $i_\1$ has already been set to $1$.  We denote the measure defined by $\hat \sigma_\ta$ elements in \eqref{sigma-hat} as $\CZ_n^{(\tq)}$ and define Renyi multi-entropy $S_n^{(\tq)}$ as
\begin{align}\label{q-renyi}
    S^{(\tq)}_n\equiv \frac{1}{1-n} {\rm Log}\,(\CZ_n^{(\tq)}/(\CZ_1^{(\tq)})^{n^{\tq-1}}).
\end{align}
Here the factor $(\CZ_1^{(\tq)})^{n^{\tq-1}}$ serves to normalize the state.
This is the family of $\tq$-partite measures that we are interested in. Let us pause for a moment to consider properties of the set $\{\hat \sigma_\ta:\ta=\2,\ldots,\tq\}$. 
\begin{itemize}
    \item $\hat \sigma_\ta$'s have the same equivalence class given by the cycle $p_n=n^{\tq-2}$ with all other $p_k=0$. 
    \item $\hat \sigma_\ta^{-1}\hat \sigma_\tb$ for $\ta\neq \tb$ also have the same equivalence class.
    \item Together, $\hat \sigma_\ta$s generate the subgroup ${\mathbb Z}_n^{\otimes {\tq-1}}$ of the permutation group. 
\end{itemize}
These observations will be important in the future discussions. 

\begin{figure}[t]
    \begin{center}
    \includegraphics[scale=.2]{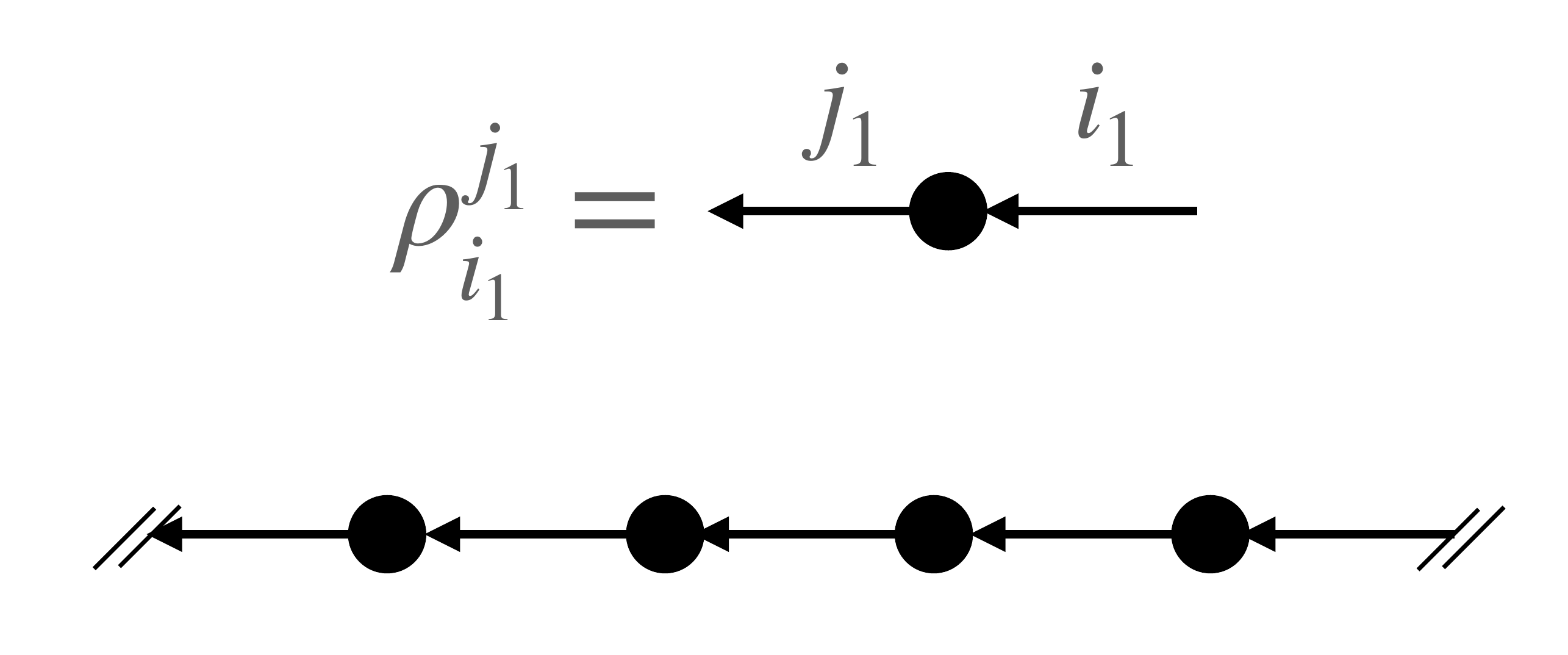}
    \end{center}
    \caption{Graphical notation for the density matrix for $\tq=2$ and the index contraction that gives $\CZ_4^{(2)}$.}\label{rho2}
\end{figure}

In a way, fixing $\hat \sigma_\1={\rm id}$ has a very natural interpretation. If we consider a single $\psi$ and a single $\bar \psi$ with $i_\1$ index contraction, we get the density matrix on $\cH_\2\otimes\ldots\otimes \cH_\tq$. This density matrix has the index structure
\begin{align}
    \rho_{\alpha_\2^{(i_\2)}\ldots\alpha_\tq^{(i_\tq)}}^{\alpha_\2^{(\hat \sigma_\2\cdot i_\2)}\ldots\alpha_\tq^{(\hat \sigma_\tq\cdot i_\tq)}}.
\end{align}
This means the gauge fixed permutations $\hat \sigma_\ta$'s describe index contractions of this density matrix. Let us see this with two examples, for $\tq=2$,
\begin{align}\notag
    \CZ^{(\2)}_n=\sum\prod_{i_2} \,\rho_{\alpha_\2^{(i_\2)}}^{\alpha_\2^{(\hat \sigma_\2\cdot i_\2)}}=\sum\rho_{\alpha_\2^{(1)}}^{\alpha_\2^{(\hat \sigma_\2\cdot 1)}} \rho_{\alpha_\2^{(2)}}^{\alpha_\2^{(\hat \sigma_\2\cdot 2)}} \ldots \rho_{\alpha_\2^{(n)}}^{\alpha_\2^{(\hat \sigma_\2\cdot n)}}.
\end{align}
Here sum is over all repeated indices.
With $\hat \sigma_\2$ being a cyclic element as specified by equation \eqref{sigma-hat}, this is nothing but $\CZ_n$ defined earlier. So we have showed $S_n^{(\2)}=S_n$. Graphically, denoting $\rho$ as in figure \ref{rho2}, we get the graphical representation of $S^{(\2)}_n$ as a circular lattice of $\rho$'s of length $n$ as in figure \ref{rho2}. Each directed link in this lattice represents index contraction from a fundamental index to an anti-fundamental index. The advantage of this graphical notation is that it admits straightforward extension to higher number of parties.
\begin{figure}[t]
    \begin{center}
    \includegraphics[scale=.2]{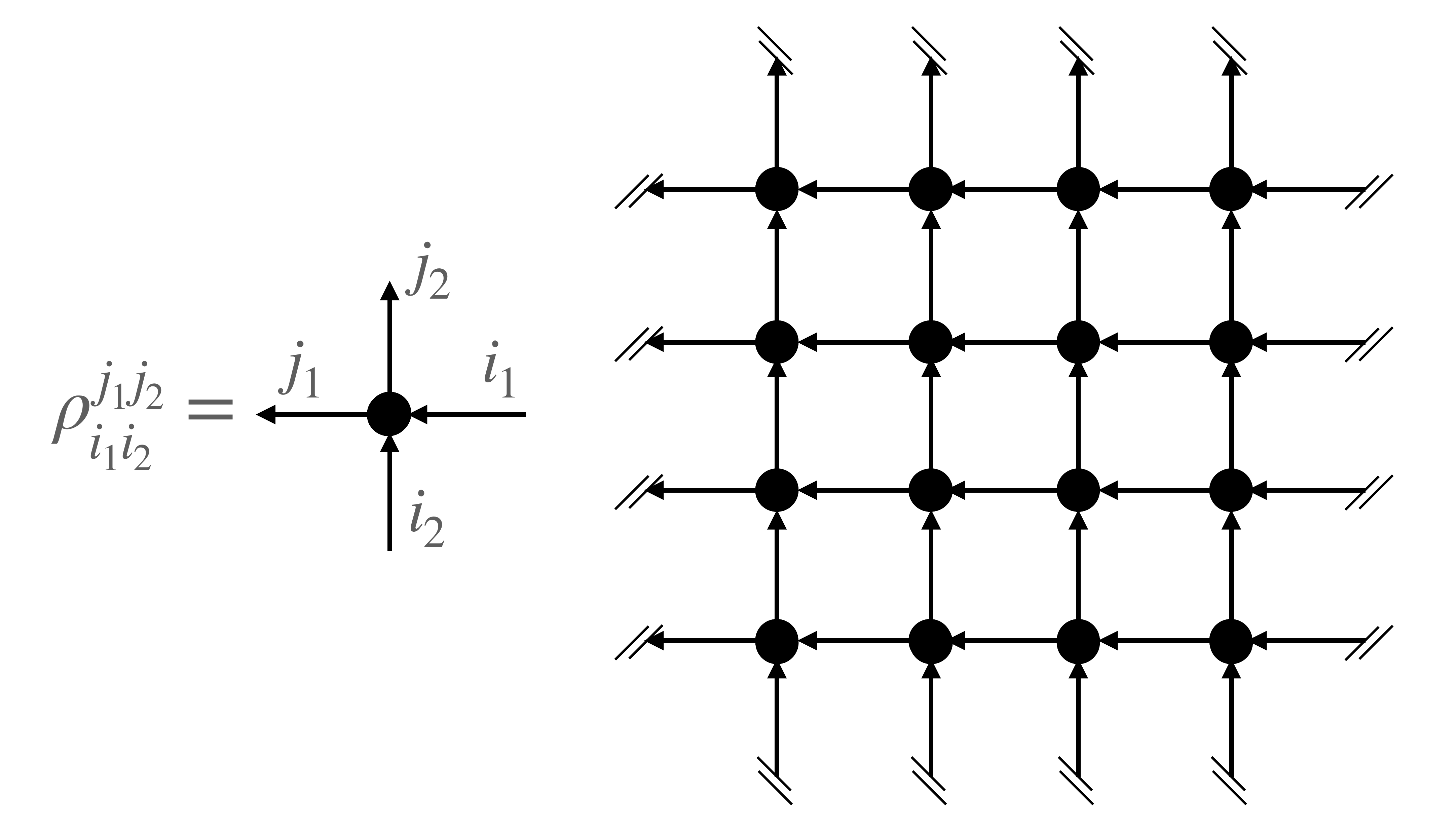}
    \end{center}
    \caption{Graphical notation for the density matrix for $\tq=3$ and the index contraction that gives $\CZ_4^{(3)}$}\label{rho3}
\end{figure}
For instance, in the case of $\3$-parties, we denote the density matrix as in figure \ref{rho3}. Then $\CZ^{(\3)}_n$ is the  index contraction given by the toric lattice of $\rho$'s with both sides being of length $n$. This is shown in figure \ref{rho3}. Similarly, $\CZ^{(\tq)}_n$ is the index contraction given by  $(\tq-1)$-dimensional toric lattice with all sides being of length $n$.

Arguably the most important measure of bi-partite entanglement, the von-Neumann entropy $S$, is obtained as a limit
\begin{align}
    S={\rm lim}_{n\to \1}\, S_n.
\end{align}
The Renyi multi-entropy $S^{(\tq)}_n$ is defined such that the limit
\begin{align}\label{limitq}
    S^{(\tq)}={\rm lim}_{n\to \1}\,  S^{(\tq)}_n,
\end{align}
called \emph{the} multi-entropy, has a number of nice properties.
\begin{enumerate}
    \item It is symmetric in all the parties.
    \item For $\tq=2$, it reduces to the von-Neumann entropy.
    \item It admits a convenient holographic description.
\end{enumerate}
Properties $1$ and $2$ are obvious from the discussion so far.  In the rest of the paper, we will discuss multi-entropy for holographic theories and describe what we exactly mean by property $3$. The multi-entropy can also be computed from ${\CZ}_n^{(\tq)}$ using the formula
\begin{align}
    S^{(\tq)}=-\partial_n {\rm log}\Big(\CZ_{n}^{(\tq)}/(\CZ_{1}^{(\tq )})^{n^{\tq-1}}\Big)|_{n=1}.
\end{align}
We will use this formula and equation \eqref{limitq} interchangeably.

\subsection{A note on analytic continuation}
We have defined the multi-entropy as a limit of the Renyi multi-entropy in equation \eqref{limitq}. This requires analytic continuation of $S_n^{(\tq)}$ away from integer $n$. One may rightly wonder whether such an analytic continuation exists and under what conditions is it unique. The issue of existence was recently emphasized in \cite{Pen:2022dhr}. 
The uniqueness issue is usually tackled using the Carlson's theorem. It says
\begin{itemize}
    \item A function $f(z)$ that is analytic for ${\rm Re}(z)>0$ and satisfies
    \begin{align}
        f(|z|)\leq A e^{c |z|},\quad f(iy)\leq A e^{\pi y} 
    \end{align}
    for some real constants $A$ and $c$ for all $z\in {\mathbb C}$ and $y\in {\mathbb R}$ and vanishes for any non-negative integer must be identically zero.
\end{itemize}
In order to argue for the uniqueness of the analytic continuation (if it exists) we must look for analytic functions obeying the analyticity and boundedness conditions required by the Carlson's theorem, henceforth called the Carlson conditions. The problem of constructing such a function from its values at non-negative integers or ``input data'' was considered in a paper by Regge and Viano \cite{regge-viano}. There this problem was motivated from the theory of complex angular momentum. The authors obtain a solution that is even amenable to numerical analysis. However, it is not clear whether their  proposed analytic continuation obeys the Carlson conditions. This question was expounded upon in the paper \cite{osti_4065624}. In this paper, authors propose an elaborate set of conditions on the input data that is necessary for the Carlson conditions to hold \footnote{We would like to thank Geoff Penington to bring this paper to our attention.}. We are currently exploring if these conditions satisfied Renyi multi-entropies of qubit states. If these conditions hold for general quantum states then the multi-entropy would be unambiguously defined. For the purposes of this paper, we will define multi-entropy only for those states whose Renyi multi-entropy admit a manifest analytic continuation. As we will see below, this is the case for holographic states.

\subsection{Example}
Although the multi-entropy is amenable to holographic computation as we will see shortly, it is difficult to compute
for finite dimensional quantum systems. This is because multi-entropy is defined only via the replica trick and, in general, it is difficult to compute $\CZ_n^{(\tq)}$ as an analytic function of $n$. However, for a special class of states that we call generalized GHZ states, the multi-entropy is readily computed. Consider $|\Psi\rangle \in \otimes_{\ta=1}^\tq \CH_\ta$ with ${\rm dim}\, \CH_{\ta}=d$. The generalized GHZ state is defined as 
\begin{align}
    |\Psi\rangle_{\rm GHZ}=\sum_i \lambda_i |i\rangle\otimes\ldots\otimes |i\rangle.
\end{align}
Here $|i\rangle$ is an orthonormal basis in $\CH_{\ta}$ for all $\ta$. The state is normalized so $\sum_i |\lambda_i|^2=1$.
It is easy to compute $\CZ_n^{(\tq)}$ because the same index $i$ runs through all the contractions. It contributes $|\lambda_i|^{2n^{\tq-1}}$. Then the $\tq$-Renyi entropy is 
\begin{align}
    S_n^{(\tq)}=\frac{1}{1-n}{\rm log}\Big(\sum_i |\lambda_i|^{2n^{\tq-1}} \Big).
\end{align}
Taking the $n\to 1$ limit,
\begin{align}
    S^{(\tq)}= (1-\tq)\sum_i|\lambda_i|^2{\rm log}|\lambda_i|^2.
\end{align}
A bi-partite state can always be taken to the generalized GHZ form via Schmidt decomposition. Then $|\lambda_i|^2$ are the eigenvalues of the density matrix. It is clear that $S^{(\tq)}$ agrees with the von-Neumann entropy for $\tq=2$. We believe that it is extremely important to develop techniques to compute multi-entropy for general states to understand its  quantum information theoretic properties.

\subsection{Distinguishing isospectral density matrices}

In this section, we will discuss the effectiveness of Renyi multi-entropy in distinguishing multi-partite entanglement in two mixed states. For this purpose, we will present density matrices $\rho_{12}$ and $\sigma_{12}$  on two qubits such that the spectrum of $\rho_{12}, \rho_1$ and $\rho_2$ is identical to that of $\sigma_{12}, \sigma_1$ and $\sigma_2$ respectively. Such density matrices are known as isospectral. This example of the isospectral pair is borrowed from \cite{nielsen2001separable}. The isospectral property of these density matrices guarantees that no bi-partite measure of entanglement can distinguish between these states. However, as we will show, all the Renyi multi-entropies for these two states are different. 
\begin{align}
    \rho_{12}=\frac13
    \begin{pmatrix}
        1 & 0 & 0 & 0\\
        0 & 1 & 1 & 0\\
        0 & 1 & 1 & 0\\
        0 & 0 & 0 & 0
    \end{pmatrix},\quad 
    \sigma_{12}=\frac13
    \begin{pmatrix}
        1 & 0 & 0 & 0\\
        0 & 0 & 0 & 0\\
        0 & 0 & 0 & 0\\
        0 & 0 & 0 & 2
    \end{pmatrix}
\end{align}
The non-zero eigenvalues of $\rho_{12},\rho_1,\rho_2$ (and also for $\sigma_{12},\sigma_1,\sigma_2$) are $\frac13,\frac23$. The values of the first few Renyi multi-entropies are
\begin{center}
    {\renewcommand{\arraystretch}{1.8}
    \begin{tabular}{ c| c | c }
      &$\rho$ & $\sigma$ \\ 
     \hline
     $S^{(3)}_2$  & ${\rm Log}\left(9\right)$ & ${\rm Log}\left(\frac{81}{17}\right)$ \\  
     \hline
     $S^{(3)}_3$ & $\frac12{\rm Log}\left(\frac{6561}{14}\right)$ & $\frac12{\rm Log}\left(\frac{729}{19}\right)$  \\  
     \hline
     $S^{(3)}_4$ & $\frac13{\rm Log}\left(\frac{14348907}{139}\right)$ & $\frac13{\rm Log}\left(\frac{43046721}{65537}\right)$ 
    \end{tabular}}
\end{center}

In fact, it is easy to see that $\rho_{12}$ is obtained by tracing out one party in the pure W-state $(|100\rangle+|010\rangle+|001\rangle)/\sqrt{3}$ and $\sigma_{12}$ is obtained analogously from the pure generalized GHZ state $(|000\rangle+\sqrt{2}|111\rangle)/\sqrt{3}$. The density matrix $\sigma_{12}$ is separable while the density matrix $\rho_{12}$ is not. The fact that the (Renyi) multi-entropy distinguishes the two suggests that it can lead to a new separability criterion. We are currently exploring this possibility.

\section{Multi-entropy from Holography}\label{holography}
In this section, we will discuss computation of  multi-entropy in a $D$-dimensional conformal field theory ${\cal T}$. Let the state  $|\Psi\rangle$ be defined on a time symmetric Cauchy slice $\CR$ of a $D$-manifold $\CM$. It is given by a Euclidean path integral on the half-space $\CM_\Psi$ such that $\partial \CM_\Psi=\CR$.  The dual  bra  $\langle \Psi|$ is constructed by Euclidean path integral on the other half $\CM_{\bar \Psi}$. It is obtained from $\CM_\Psi$ by reflecting across $\CR$. The squared norm of $\Psi$  is the partition function $\CZ_\CM$  on $\CM$. 
Let us decompose $\CR$ into  $\tq$ number of disjoint regions $\CR_\ta$,  such that $\cup_\ta \CR_\ta=\CR$. Let the Hilbert space on region $\CR_\ta$ be $\cH_\ta$.
We are interested in computing multi-entropy of the state $|\Psi\rangle$ under the decomposition  $\otimes_\ta \cH_\ta$. For theories that admit a weakly coupled gravity dual, this problem can be addressed holographically.

The replica trick \cite{Calabrese:2004eu}  involves working with the tensor product theory ${\cal T}^{\otimes n^{\tq-1}}$ on $\CM$. This theory has the discrete symmetry ${\mathbb S}_{n^{\tq-1}}$. It admits co-dimension $2$ twist defects labeled by elements of this permutation group. 
For every pair of regions $(\CR_\ta,\CR_\tb)$ that share a boundary, we insert the twist operator ${\CO}_{\hat \sigma_\ta^{-1}\hat \sigma_{\tb}}$ on the common boundary. The measure $\CZ_n^{(\tq)}$ is then given by the correlation function of these twist operators.
Let us denote the resulting replicated manifold as ${\cal M}_n$. The correlation function of twist operators is the partition function $\CZ_{\CM_n}$ on $\CM_n$.
Using \eqref{q-renyi} we have
\begin{align}\label{q-replica}
    S^{\tq}_n=\frac{1}{1-n}{\rm Log}(\CZ_{\CM_n}/(\CZ_\CM)^{n^{\tq-1}}).
\end{align}

Following \cite{Lewkowycz:2013nqa}, we will proceed to analyze this problem holographically. Let $\CB_n$ be dominant the gravity solutions such that $\partial \CB_n=\CM_n$. The manifold $\CB_n$ is a smooth manifold e.g. in the case of Einstein gravity with negative cosmological constant, it is of constant negative curvature. The holographic dictionary gives
\begin{align}\label{holo-partition}
   {\rm Log} \,\CZ_{\CM_n}=-\CS_{\rm grav}(\CB_n).
\end{align}
Here $\CS_{\rm grav}({\cal X})$ is the gravitational action evaluated on the solution ${\cal X}$.
For $n=1$, this gives ${\rm Log}\, \CZ_{\CM}=-\CS_{\rm grav} (\CB), \CB\equiv \CB_1$.  

The background fields on the manifold $\CM_n$ enjoy a replica symmetry. This is the symmetry generated by the permutation elements $\hat \sigma_\ta^{-1}\hat \sigma_\tb$ associated to all the twist operators. As remarked earlier, this group is ${\mathbb Z}_n^{\otimes {\tq-1}}$. Following \cite{Lewkowycz:2013nqa}, we will assume that the dominant bulk solution $\CB_n$ enjoys this symmetry. 
The solution $\CB_n$ consists of co-dimension $2$ loci that are invariant under certain subgroups of the replica symmetry group.  Some of these loci, called ``external'', are anchored at the fixed points on the boundary (these are locations of twist operator insertions on $\CM$) while the rest are ``internal''. Let us denote the loci that are anchored at the fixed points corresponding to the twist operator $\CO_g$ as $\CL_g$. They are invariant under the ${\mathbb Z}_n$ subgroup generated by $g$. Generically, two $\CL$s can merge to form a different $\CL$. Merging obeys the algebra $\CL_{g_1}\cdot \CL_{g_2}\to \CL_{g_1g_2}$.
The internal loci come about because of such merging. 
Below we will assume that 
\begin{enumerate}
    \item Every fixed point locus is of the form $\CL_{\hat \sigma_\ta^{-1}\hat \sigma_\tb}$.\label{assume-1}
\end{enumerate}
We will justify this assumption shortly. Thanks to the special property of our permutation elements $\hat \sigma_\ta$ stated below equation \eqref{q-renyi},  this means that any locus is invariant under \emph{some} ${\mathbb Z}_n$ subgroup of the replica symmetry. Every locus $\CL_g$ appears in groups of $n^{\tq-2}$. This is because, the orbit of the action of replica group ${\mathbb Z}_n^{\otimes {\tq-1}}$ consists of $n^{\tq-2}$ elements as its stabilizer is ${\mathbb Z}_n$.  For $\tq=2$, the number of elements in the orbit is $1$.

We now make use of the replica symmetry in the bulk to construct the orbifold $\widetilde \CB_n\equiv \CB_n/({\mathbb Z}_n)^{{\tq-1}}$. Due to symmetry, the classical gravitational actions on the two spaces are related as
\begin{align}\label{orbifold-action}
    \CS_{\rm grav}(\CB_n)=n^{\tq-1}\,\CS_{\rm grav}(\widetilde \CB_n).
\end{align} 
The orbifold $\widetilde \CB_n$ has a nice  property that $\partial \widetilde \CB_n=\CM$. A group of $n^{\tq-2}$ number of  $\CL_g$ become a single conical singularity of opening angle $2\pi/n$ in the orbifold $\widetilde \CB_n$. Let us denote this singularity as $\widetilde \CL_g$. 
Let us denote the web created by these singularities as $\CW$.  
Consider a co-dimension $1$ slice $\CC\in \widetilde \CB_n$ that contains $\CW$ and $\partial \CC=\CR$. There are multiple such slices and the precise choice doesn't matter for the following discussion. Every singularity becomes a co-dimension $1$ wall in $\CC$ and the web $\CW$ yields its chamber decomposition. As we move from $\CR_\ta$ to $\CR_\tb$ through $\CC$, we must encounter at least one wall because the permutation elements $\hat \sigma_\ta$ and $\hat \sigma_\tb$ are different.
At this stage, we make an  assumption about $\CW$  that
\begin{enumerate}
    \setcounter{enumi}{1}
    \item There is no chamber which lies  completely in the interior of $\CC$.\label{assume-2}
\end{enumerate}
As a result, we get a one-to-one map between the chambers and boundary regions. Let us denote the chamber adjacent to $\CR_\ta$ as $\CC_\ta$. It has the property $\partial\CC_a\cap \CM=\CR_\ta$. The web $\CW$ consists of only those walls that separate $\CC_\ta$ and $\CC_\tb$ for some $(\ta,\tb)$. Such a wall must be of the type $\widetilde \CL_{\hat \sigma_\ta^{-1}\hat \sigma_\tb}$. The parent $\CL\in \CB_n$ must also be of the same type. This justifies our assumption \ref{assume-1}. 

To compute the $\tq$-Renyi entropy using equations \eqref{holo-partition} and \eqref{orbifold-action}, we need to evaluate the gravitational action on the orbifold solution $\widetilde{\CB}_n$. In what follows, we will specialize to the case of Einstein gravity. The orbifold solution is a smooth solution of Einstein's equations with constant negative curvature except at  $\widetilde{\CL}$ where it has a conical singularity of opening angle $2\pi/n$. The gravitational action for this solution is computed as follows \cite{Lewkowycz:2013nqa}. We excise a small tubular neighborhood, say of radius $a$, around the conical singularity ${\widetilde \CL}$. The gravitational action comes from the Gibbons-Hawking-York (GHY) term evaluated on the resulting boundary. This contribution is extensive in the co-dimension $2$ area of ${\widetilde \CL}$. In the limit $a\to 0$, the action takes the form \cite{Dong:2016fnf},  $\frac{1-n}{4nG_{N}}\int dy^{D-1} \sqrt{h}$. Here $y$ is a coordinate along ${\widetilde \CL}$ and $h$ is the induced metric on ${\widetilde \CL}$. To compute the equations that are obeyed by the singular locus, it is useful to introduce a cosmic brane with the same action supported on the singularity.
\begin{align}\label{Sbr}
    \CS_{\rm br}^{(n)}=\frac{1-n}{4nG_{N}}\int dy^{D-1} \sqrt{h}=\frac{1-n}{4nG_{N}}A. 
\end{align}
The $y$ integral is simply the area $A$ of the brane web $\CW$. The solution is then computed by solving the equations coming from the action ${\cal S}_{\rm grav}+{\cal S}^{(n)}_{\rm br}$. It is known \cite{Unruh:1989hy, Boisseau:1996bp} that such a cosmic brane action indeed gives rise to a conical singularity of opening angle $2\pi/n$ in the two transverse directions as desired. Once the solution is found by extremizing $\CS_{\rm grav}+\CS_{\rm br}^{(n)}$, we need to evaluate only the action $S_{\rm grav}$ on the solution. This is because the cosmic brane is not actually present at the singularity but merely used as a trick to model the singularity. 
The novelty in the multi-partite case is that multiple  $\widetilde{\CL}_g$'s can meet each other at higher co-dimensional loci. To accommodate such a meeting we simply let the corresponding cosmic branes meet. 

\begin{figure}[t]
    \begin{center}
    \includegraphics[scale=.2]{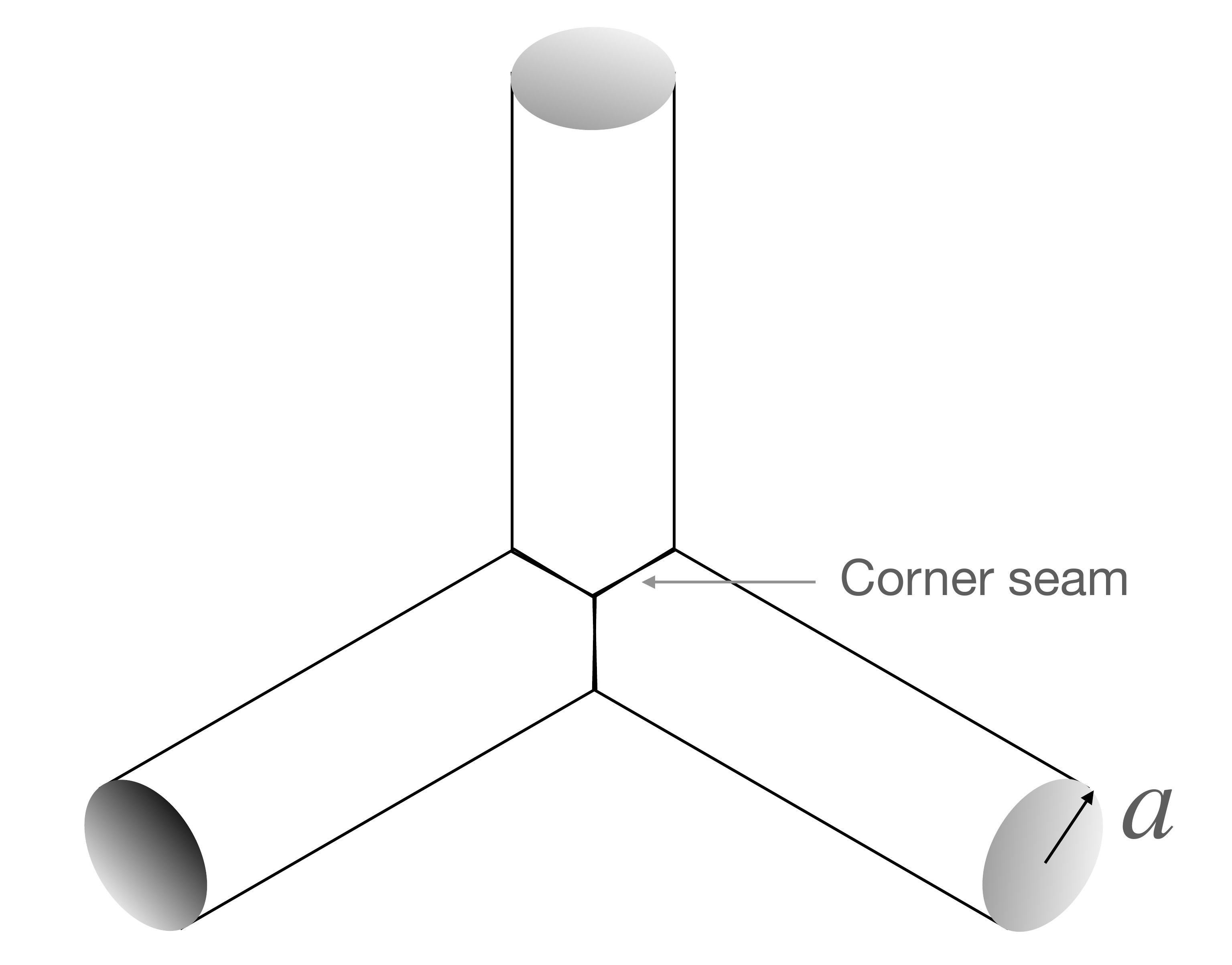}
    \end{center}
    \caption{Top view of the junction of  tubular neighborhoods of three ${\widetilde \CL}'s$.}\label{corner}
\end{figure}

A priori, there can be additional terms in the brane action supported only at the  meeting locus. To compute such terms we again excise the tubular neighborhood of ${\widetilde \CL}$'s and focus near their higher co-dimensional junction. As multiple ${\widetilde \CL}$'s meet, their corresponding tubular neighborhoods also meet forming corners as shown in figure \ref{corner}. 
For the gravitational variational principle to be well-defined, we need to add the so-called Hayward term at the corner just as the way we add the GHY term on the boundary.
\begin{align}
    {\cal S}_{\rm Hayward}=-\frac{1}{8\pi G_N} \int d\xi^{D-2}\, (\theta-\pi) \sqrt {\gamma}. 
\end{align}
Here $\theta$ is the angle subtended by the two boundaries at the corner, $\xi$ is the coordinate along the corner and $\gamma$ is the induced metric on the corner. The contribution of the Hayward term is extensive along the co-dimension $3$ meeting locus and the proportionality constant is computed by integrating the Hayward term  along the one-dimensional corner seam of the tubular neighborhood as shown in figure \ref{corner}. This integral is proportional to $a$ and vanishes as we take $a\to 0$. This had to be the case on dimensional grounds because the contribution at the meeting locus must take the form  $\sim \frac{1}{ G_N} \ell^{(3)}_{\rm meeting} \ell_{\rm scale}$.
Here $\ell^{(3)}_{\rm meeting}$ is the length (co-dimension $3$) of the meeting locus in $\CW$ and $\ell_{\rm scale}$ is some length scale needed to obtain a dimensionless answer. The only length scale that could serve this purpose is $a$ \footnote{Another length scale present in the problem is the AdS scale $\ell_{\rm AdS}$, however it is irrelevant to this completely local computation.} which we take to $0$. This argument also shows that there is no extra contribution to the action even at higher co-dimension meeting loci.
 
Finding the solution to the theory $\CS_{\rm grav}+\CS_{\rm br}^{(n)}$ for general $n$ is still a daunting task (see \cite{Headrick:2010zt, Hung:2011nu, Dong:2016fnf} for computation of bi-partite Renyi entropy). But to compute multi-entropy, only the limit $n\to 1$ is relevant. As the only $n$ dependence appears in the coefficient in $\CS_{\rm br}^{(n)}$, the solution can be analytically continued away from $n$ integer. Moreover, in the limit $n\to 1$, the tension of the brane goes to zero and the solution can be found in the probe limit as brane web configuration that extremizes $\CS_{\rm br}$ in the fixed background $\CB$. The solution obeys the equation of motion $ \delta_{g_{\mu\nu}} (\CS_{\rm grav}+\CS_{\rm br}^{(n)})=0$. In particular,
\begin{align}
    \partial_n\CS_{\rm grav}=-\partial_n\CS_{\rm br}^{(n)}=\frac{A}{4G_N}.
\end{align}
This, along with equations \eqref{limitq}, \eqref{q-replica} and \eqref{holo-partition}, shows that the multi-entropy is given a simple formula
\begin{align}
    S^{(\tq)}=\frac{A(\CW)}{4G_N}.
\end{align}
Here $A(\CW)$ is the area of the minimal brane web $\CW$ in $\CB$. 
The brane web $\CW$ obeys the topological conditions,
\begin{enumerate}
    \item $\CW$ is anchored at the boundaries of all the regions $\CR_\ta$'s.
    \item $\CW$ contains sub-webs that are homologous to all the regions $\CR_\ta$'s.
\end{enumerate} 
The second condition is the reformulation of the statement that between any two chambers $\CC_\ta$ and $\CC_\tb$ there must be at least one wall. As the solution minimizes the area subject to these conditions, it doesn't allow any chamber that lies completely in the interior of $\CC$. This justifies our assumption \ref{assume-2}.

The brane web in question is extremely familiar in $D=4$. In the three dimensional Cauchy slice $\CC$ it resembles a soap film in hyperbolic space anchored on a given ``wire frame'' at infinity. Specifying the anchor does not specify the soap film uniquely but when combined with the homology condition and the global minimum condition, it does. The close analogy to soap-films leads us to call our prescription, the soap-film prescription. To describe the soap-film it is convenient to introduce the label ${\widetilde \CL}^{(k)}$ for the special locus of co-dimension $k$. It is defined inductively as the meeting locus of ${\widetilde \CL}^{(k-1)}$ with ${\widetilde \CL}^{(2)}\equiv {\widetilde \CL}$. For $D=4$ it is known that
\begin{itemize}
    \item Three ${\widetilde \CL}^{(2)}$'s meet at ${\widetilde \CL}^{(3)}$ at an angle $2\pi/3=\cos^{-1}(-1/2)$.
    \item Four ${\widetilde \CL}^{(3)}$'s meet at ${\widetilde \CL}^{(4)}$ at an angle $\cos^{-1}(-1/3)$.
\end{itemize}
These are known as Plateau's laws of soap-film. In dimension $D$ we expect $k+1$ of ${\widetilde \CL}^{(k)}$'s to meet 
at ${\widetilde \CL}^{(k+1)}$ at an angle $\cos^{-1}(-1/k)$ and so on until we get to ${\widetilde \CL}^{(D)}$. 

We have verified our proposal for $2D$ CFTs with large central charge. These calculations will appear in an accompanying paper \cite{multi-long}.

\section{Generalizations}

In this section we will comment on the generalizations of our soap-film prescription in various directions.

\subsection{Covariant prescription}

The covariant generalization of the Ryu-Takayanagi formula was presented in \cite{Hubeny:2007xt}. It was argued there that the entanglement entropy is given by area of the extremal surface rather than the minimal one. This prescription was later rephrased as a maximin prescription in \cite{Wall_2014}, where the surface in question is obtained first by minimizing the area on some achronal slice $\Sigma$ and then by maximizing the area with respect to the variation of $\Sigma$. In case there are multiple extremal solutions, we pick the one with the minimum area as that corresponds to the most dominant solution.

The soap-film prescription proposed here admits a natural covariant generalization along the lines of \cite{Hubeny:2007xt} and \cite{Wall_2014}. We conjecture that the covariant multi-entropy is obtained as the area of the minimal extremal soap-film rather than the globally minimal one and that this choice is equivalent to the one obtained by the maximin prescription of \cite{Wall_2014}. 

\subsection{Higher derivative gravity}

For higher derivative gravitational theories, we do expect a non-trivial contribution localized at the higher co-dimensional meeting loci ${\widetilde \CL}^{(k)}$'s because the dimensional analysis argument presented in section \ref{holography} ceases to be  valid as $\ell_{\rm scale}$ can be provided by the inverse mass-scale of the higher derivative corrections. 
It would be interesting to compute these terms in general higher derivative theory of gravity.

\subsection{Quantum corrections}
Quantum correction to the multi-entropy can be found  using the replica trick in the bulk.
We consider $n^{\tq-1}$ replica copies of the orbifold $\widetilde \CB_n$ and insert the bulk twist operators $\CV_g$ at $\widetilde \CL_g$. This has the effect of reversing the orbifold and give back the geometry $\CB_n$. The partition function on the replicated manifold is then simply the partition function  $\CZ_n^{(\tq)}$. This is the partition function that goes into the calculation of the multi-entropy. If the replica trick was performed on the bulk geometry $\CB$, it would have given the bulk multi-entropy $S^{(\tq)}_{\rm bulk}(\CW)$ corresponding to the chamber decomposition $\CC_a$ directly, however, because the replica trick was performed on $\widetilde \CB_n$, it is not obvious that what we get is $S^{(\tq)}_{\rm bulk}(\CW)$. 
This situation is similar to the bi-partite case \cite{Faulkner:2013ana} (see also \cite{Barrella:2013wja}). There, the difference between the two quantities is captured by changing the classical solution by $\CO(G_N)$ to account for the one loop expectation value of the stress tensor. This changes the area and hence the entanglement entropy by $\CO(1)$. We expect a similar formula to give the sub-leading correction to the multi-entropy 
\begin{align}\label{quantum-multi}
    S^{(\tq)}=\frac{\langle\hat A(\CW)\rangle}{4G_N}+S_{\rm bulk}^{(\tq)}(\CW)+{\rm c.t.}.
\end{align}
Here $\hat A(\CW)$ is the area operator of the soap film and ${\rm c.t.}$ are the counter-terms that render $S_{\rm bulk}^{(\tq)}(\CW)$ finite. 

Following \cite{Engelhardt:2014gca}, \cite{Dong:2017xht} we conjecture a formula that is valid to all orders in $1/G_N$ perturbation theory: Multi-entropy is given 
by the above formula but $\CW$ is not the ordinary area extremizing soap-film but rather the ``quantum extremal soap-film'' i.e. the soap film that extremizes the combination 
\begin{align}
    \frac{ A(\CW)}{4G_N}+S_{\rm bulk}^{(\tq)}(\CW).
\end{align}
In analogy with the bi-partite case, we call this prescription the quantum extremal soap-film prescription.

\section*{Acknowledgements}
We would like to thank  Akshay Gadde, Gautam Mandal, Shiraz Minwalla, Piyush Shrivastava, Pushkal Shrivastava and Sandip Trivedi for interesting discussions. We are particularly indebted to Onkar Parrikar for stimulating discussions and insightful comments. We would also like to thank Shraiyance Jain for collaboration on related projects.
This work is supported by the Infosys Endowment for the study of the Quantum Structure of Spacetime and by the SERB Ramanujan fellowship.  We acknowledge support of the Department of Atomic Energy, Government of India, under Project Identification No. RTI 4002. We would also like to acknowledge our debt to the people of India for their steady support to the study of the basic sciences.

\bibliography{LargeDCFT}

\end{document}